\documentstyle[preprint,aps,epsf]{revtex}                  
\input epsf.tex
\def\DESepsf(#1 width #2){\epsfxsize=#2 \epsfbox{#1}}
\global\let\epsfloaded=Y 
\begin{document}
\pagestyle{empty}                                      
\preprint{
\font\fortssbx=cmssbx10 scaled \magstep2
\hbox to \hsize{
\hfill $
\vtop{
 \hbox{ }}$
}
}
\draft
\vfill
\title{Future Island Universes in a Background
Universe Accelerated \\
by Cosmological Constant and by Quintessence}
\vfill
\author{Tzihong Chiueh and Xiao-Gang He}
\address{
\rm Department of Physics, National Taiwan University,
Taipei, Taiwan 10764, R.O.C.}

%
%
\vfill
\maketitle
\begin{abstract}
We study bound object formation in a background universe 
accelerated by cosmological constant and by quintessence.
If the acceleration lasts forever, due to the existence of 
event horizon, one would have naively expected the universe 
to approach a state of cold death. However, 
we find that many local regions in the universe can in fact be 
protected by their own gravity to form mini-universes,
provided that their present matter densities exceed some 
critical value.  
In the case with 
cosmological constant ($\Lambda$CDM cosmology) the condition of 
forming mini-universe is that the ratio of present density parameters 
$\Omega_{i}^0/\Omega_\Lambda$ ought to be larger than a critical 
value $3.63$. Such mini-universes typically
weigh less than $2\times 10^{14}$ solar masses, with
the lighter ones having tight and compact configurations. 
In the case with quintessence, the final ratio $\Omega_i/\Omega_q$ of 
a mini-universe is found to be always larger than $w_q^2-w_q$, where 
$w_q$ is the equation of state parameter.  
\end{abstract}
%
%
%
%
\pagestyle{plain}

\section{Introduction}

Recently direct evidences from the studies of Hubble diagram
for Type Ia supernovae\cite{1} indicate that our universe is expanding
with an increasing rate, i.e., accelerating expansion.  
This result compels one to
seriously consider a dominant component of exotic dark energy, 
the quintessence, 
in the cosmic energy balance.
Accelerating expansion implies that the deceleration parameter
$q=(\Omega_m+(3w_q+1)\Omega_q)/2$
to be negative, where $w_q$ is given by the cosmic equation of
state, $p_q=w_q\rho_q$, of the quintessence $Q$.
As long as $w_q<-1/3$, the corresponding dark energy provides a 
negative contribution to $q$, and   
the relevant range of $w_q$ for acceleration 
lies in between $-1/3$ and $-1$.
When $w_q$ assumes the extreme value $-1$, 
constrained by the weak energy condition\cite{2}, the dark energy is 
reduced to the one arising from the cosmological constant $\Lambda$. 

The acceleration may stop in the future, i.e., a transient 
phenomenon with a time-dependent $w_q$, or may last forever,
depending on the nature of the dark energy\cite{3}.
A positive cosmological constant, consistent with
the supernovae data at $z=1.7$\cite{4}, leads to a forever acceleration.
Quintessence scenario with a constant $w_q$, not ruled out by data, can
also lead to forever acceleration.
There are profound implications for a forever accelerating universe\cite{5,6}.
The universe will exhibit an event horizon; that is, there exist regions of the 
universe inaccessible to light probes. 
It has been argued that such a universe
presents a challenge for string theories or any of
its alternatives\cite{5}.
Many other implications of cosmological constant and quintessence and specific
models have been studied in the literature\cite{aa,bb,cc}.

Naively speaking,
in a forever accelerating universe,
two points at different spatial locations will, as measured at any one of the two points, 
eventually be pulled apart at
the speed of light to approach the future horizon.  Hence existing 
structures will
be eternally frozen and the universe will approach a state of cold death.
Is this an unavoidable consequence everywhere in the universe, or may some
mechanisms be in operation that salvage the future?
The present work confronts this issue
and shows that the self-gravity of matter can protect a local universe
from ending up with cold death, provided that the present local matter density
is sufficiently high.
Contrary to the above naive expectation, a large number of mini-universes, 
including our own Local Group, will become self-bound and
survive the destruction
of cosmic repulsive force due to the quintessence.
They can naturally arise despite that
these mini-universes will still be falling into each
other's horizon at a sufficiently late time and become isolated island
universes.

At present, a variety of dark energy candidates
can accommodate the experimental data. 
If the dark energy arises from a positive cosmological constant, 
the $\Lambda CDM$ cosmology, data from recent cosmic
microwave background (CMB) radiation\cite{7a} further constrain the 
present energy
density parameters to be $\Omega^0_m=0.35$ and $\Omega^0_\Lambda=0.65$.
While for quintessence, depending on the value $w_q$, the present
matter density $\Omega_m^0$ and quintessence density $\Omega^0_q$ can
be different\cite{bond}.

For our purpose of addressing the future of local bound
objects in an accelerating universe, the cosmological constant presents the
worst scenario since the cosmological constant gives rise to the strongest 
repulsive force against formation of bound objects. 
If objects are not torn up by the cosmological constant repulsive force,
they will survive in other scenarios.

The paper is arranged as the following.
In Sec.(2), we first consider the $\Lambda$CDM cosmology in details, 
by assuming that the present universe is spatially flat with
65\% energy given by the cosmological constant ($\Omega^0_\Lambda = 0.65$) and
another 35\% by the matter ($\Omega_m^0 = 0.35$). 
In Sec.(3), we then consider the case where the cosmological constant 
is replaced by the quintessence with a constant $w_q$. 
In both cases the mini-universe can form by
self-gravity, but the properties of critically bound mini-universes
are very different.  Our conclusion is given in Sec.(4).

\section{Bound Object Formation with Cosmological Constant}
 
We shall first adopt a toy model
to address the questions of whether bound objects may still
form after
the background expansion has started to accelerate and how
these objects become virialized.
We then study a more realistic model where the
observation/simulation-motivated features of virialized objects are
taken into account. 

\subsection{Self-Binding by Local Gravity}

The toy evolutionary model for bound object formation assumes the 
universe to consist of
the background and an isolated cold dark matter fluctuation that grew from a
small-amplitude since the early epoch.
The localized perturbation possesses spherical
symmetry and consists of two concentric solid spheres of
different densities.  The inner sphere has a
radius $r_i$ with a uniform dark-matter density $\rho_{i}$,
which is greater than the background matter density $\rho_m$.  The outer
sphere has a radius $r_{out}$ and a uniform compensating under-density,
such that the averaged matter density within $r_{out}$ equals $\rho_m$.
Any particle outside $r_{out}$ feels no extra gravity resulting from the
over-density and expands as it would have been in a homogeneous
universe.  The inner sphere
can be regarded as a closed Friedmann universe, which
has a Hubble parameter, $H_i$, given by
\begin{eqnarray}
H^2_i &=& \left ( {\dot{r_i}\over r_i} \right)^2
= {8\pi G\over 3} (\rho_i + \rho_\Lambda) - {\kappa_i\over r_i^2},
\end{eqnarray}
where $\rho_\Lambda = (1/8\pi G)\Lambda$ is a constant and
$\kappa_i$ is the curvature of the mini-universe.

A test particle at $r_i$ feels an effective potential $V(r_i)$,
\begin{eqnarray}
V(r_i) = - {GM_i\over r_i} - {4\pi G\over 3} \rho_\Lambda r_i^2.
\end{eqnarray}
Conservation of matter yields a constant mass $M_i$ within $r_i$
with $M_i=4\pi \rho_i(r_i) r^3_i/3$.
The potential $V(r)$ has a maximum at $r_{max}$ determined by
$V'(r_{max})= 0$ with $V(r_{max}) = -4\pi G \rho_\Lambda r^2_{max}$ and
$r_{max} =(\rho_i^0/ 2\rho_\Lambda))^{1/3} r_0$,
where the
density at present has been denoted as $\rho_i^0 \equiv\rho_i(r_0)$
with $r_0$ being the
present radius of the over-dense sphere.
When the test particle reaches $r_{max}$ with a zero velocity,
it will be marginally bound. The density $\rho_i$
corresponding to this
situation is called the critical binding density, denoted as
$\rho_{ic}^0$ at present, and
any local region with density larger than $\rho_{ic}^0$ will eventually 
turn around and collapse.

The curvature of the mini-universe
can be conveniently obtained at the turn-around, after which gravitational collapse
ensues, by setting
$\dot{r_i}^2 =0$ and equating $-\kappa_i/2$ to the potential at the 
turn-around.  That is,  
\begin{eqnarray}
\kappa_i = {8\pi G\rho_\Lambda\over 3}
( {\rho_i(r_0)\over\rho_\Lambda}
{r^3_0\over r^3_{ta}} + 1) r_{ta}^2,
\label{cur}
\end{eqnarray}
where $r_{ta}$ is the turn-around radius.
In particular for a critically bound mini-universe, we have
$r_{ta}=r_{max}$ and  
$\kappa_{ic}=\kappa_{i}(r_{max}) = 8\pi G\rho_\Lambda r^2_{max}$.

The time span needed for the mini-universe to evolve to the present 
epoch is given by
\begin{eqnarray}
t_i &=& \int^{r_0}_0 {dr_i\over r_i H_i} =
\int^{r_0}_0 {dr_i\over r_i \sqrt{(8\pi G/ 3)
(\rho_i + \rho_\Lambda) - (\kappa_i / r_i^2)}}\nonumber\\
&=& {1\over\sqrt{6\pi G\rho_\Lambda}}\int^1_0
{dy\over \sqrt{(\rho_i^0/\rho_\Lambda) + y^2
-(3\kappa_i/8\pi G\rho_\Lambda r^2_0)y^{2/3}}}.
\label{mini}
\end{eqnarray}
When $t_i$ is known, Eq.(4) can be inverted to solve for the present 
local density $\rho_i^0$.  Thus one needs to obtain information about $t_i$.
To this end we note that the mini-universe should have started to grow
from a small amplitude perturbation at the beginning of matter domination
around $z=3\times 10^{3}$. Therefore to a good approximation, the evolution
time $t_i$ is equal to the background evolution time $t_B$. 
The time $t_B$ for the background Friedmann universe to reach the present
ratio $\Omega_m^0/\Omega_\Lambda^0$ is 
\begin{eqnarray}
t_B &=& \int^{a_0}_0 {da\over \dot{a}} =
{1\over\sqrt{6\pi G\rho_\Lambda}}\int^1_0
{dy\over \sqrt{(\Omega_m^0 /\Omega_\Lambda^0)+ y^2}}=
{1\over\sqrt{6\pi G\rho_\Lambda}}
\sinh^{-1}( \sqrt{{\Omega_\Lambda^0\over \Omega_m^0}}).
\label{freid}
\end{eqnarray}
Setting $t_i=t_B$,
one can determine the present density ratio
$\rho_{i}^0/\rho_\Lambda$ for a given $\kappa_i$.

We consider three cases of interest for the discussions of the
density ratio $\rho_{i}/\rho_\Lambda$: i) the critical
mini-universe, where
the test particle reaches $r_{max}$ with zero velocity and is therefore 
marginally bound; ii) the case where the over-density is 
larger than that of case i) and the
test particle has a zero velocity and turns around at present; 
iii) the over-density is even larger so that the turn around occurred at
the time when the background matter density $\rho^{eq}_m$ 
equaled $\rho_\Lambda$, a particular epoch prior to which the
matter in-fall had been vigorous.

For case i), we have 
$\kappa_i=\kappa_{ic}=8\pi G\rho_\Lambda r_{max}^2$.
Given the background density parameters
$\Omega_m^0 = 0.35$ and $\Omega_\Lambda^0 = 0.65$, Eq.(4) along with Eq.(5)
yield the present critical density ratio 
$\rho^0_{ic}/\rho_\Lambda = 3.63$ 
($\Omega_{ic}(r_0) =2.36$) and $r_0= r_{max}/1.22$.

We note that
over-dense regions with the present density parameter larger than the
critical value $\Omega_{ic}(r_0)$ are abundant in the universe.
In fact nearly all future bound objects
will never reach $r_{max}$ with a turn-around radius
$r_{ta}$ smaller than $r_{max}$.  
Inserting $\kappa_i$ of Eq.(3) into Eq.(\ref{mini}), one
obtains their present densities $\rho_i(r_0)$ for a given ratio
$r_{ta}/r_0$.
For case ii) the mini-universe that is presently turning around, we
have $r_{ta} = r_0$, and
it is found that $\rho_i^0/\rho_\Lambda = 5.90$
with $\Omega_{i,0} = 3.83$.  Any over-dense region with a density greater
than this
value has already undergone collapse.  For case iii), 
it is found that $\rho_i^{eq}/\rho_\Lambda = 8.55$
with $\Omega_{i,eq}= 5.52$. 

Although the toy evolutionary model so far discussed 
may seem somewhat simplistic, the
model nevertheless illustrates the essential physics\cite{8} and supports
our contention. 
From the above discussions, we clearly see that local bound objects can form
against the cosmic repulsive force, as long as the
local density parameter $\Omega_i(r_0)$ at present is greater
than $2.36$. This value is at least a factor of several
less than the mean density parameter of the Local Group,
which includes the Milky Way, the Andromeda galaxy and more than a dozen
of smaller galaxies such as the Magellanic clouds.  It suggests that
the Local Group has defied the cosmic repulsive force and 
gravitational collapse is underway.

\subsection{Virialization}

Almost all bound objects, except for a set of measure-zero, 
critically bound ones
will undergo, or have already undergone, gravitational collapse.
It is instructive to further examine the issue of 
virialization after an object
collapses. The virial density found here helps us
address deeper questions concerning the internal
structures of bound
objects to be addressed below.  The virialized kinetic energy
of a test particle is given by, $<T> = -<\vec F\cdot \vec r>/2$.  For the
simplified model under consideration, $<T>$ is given by
$<T> = (GM_i/2 r_v) - (4\pi G\rho_\Lambda r^2_v/3)$,
where $r_v$ is the virialized radius. This leads to
\begin{eqnarray}
-{k_i\over 2}=<T> + <V> = -({GM_i\over 2 r_v}
+{8\pi G\over 3}\rho_\Lambda r^2_v).
\end{eqnarray}
From both energy and mass conservation together with
the expression of $k_i$ at the turn-around, eq.(\ref{cur}), with
the subscript``$0$'' replaced by ``$ta$'', one finds\cite{7}
\begin{eqnarray}
{\rho_{i,ta}\over\rho_\Lambda}(2-{r_{ta}\over r_v})+
2(1-2{r_v^2\over r_{ta}^2})=0.
\label{ratio}
\end{eqnarray}

Regions with a space curvature $\kappa_i$ infinitesimally greater than
$\kappa_{ic}$ of case i) will finally collapse in the infinite future, and
we shall also regard this situation as case i) in a broad sense. 
It is straightforward to obtain
that $r_v/r_{ta}=1/2.73$ for case i), since $\rho_{i,ta}/\rho_\Lambda=2$ 
and the corresponding $r_v/r_0$ is
$1/2.24$. The ratio of
the virialized density to
the $\Lambda$ energy density is then found to be
$\rho_v/\rho_\Lambda=41$.
For cases ii), we find $r_v/r_0 =1/2.20$
and $\rho_v/\rho_\Lambda = 62$.  For case iii), $r_v/r_{ta} = 1/2.13$ and
$\rho_v/\rho_\Lambda = 83$.

Case (iii) is interesting from the consideration that the matter in-fall
was considerably vigorous prior to the turn-around but became suppressed
afterward.
Although its collapse started 
at $z=0.23$ when $\Omega_m = \Omega_\Lambda$, as
the completion of collapse takes a time span twice the turn-around time,
this object will be virialized in the future
when the background universe reaches $\Omega_\Lambda=0.89$.  By then the
background scaling
factor $a(t)$ becomes a factor $1.63$ times the present value $a_0$.

After the bound object collapses and forms a virialized object,
what does the bound object need to adjust so as
to guard itself against the stripping repulsive force?  We shall
address this question by a model of somewhat more sophistication.
It has been empirically found by cosmological $N$-body simulations,
which compute at best up to the present epoch, that cold
dark matter particles tend to cluster into individual bound objects
with a universal density profile.
This density profile has been fitted by the empirical formula:
\begin{equation}
\rho=\frac{\Delta}{r(r_s+r)^2},
\label{univ}
\end{equation}
where $\Delta$ and $r_s$ characterize the mass and inner scale length of
the bound object\cite{9}.  This density profile has a $r^{-1}$ cusp and
a $r^{-3}$ envelope extended out to the virial radius $r_v$.  It
is universal in that the parameterized
functional form is independent of cosmological parameters and epochs,
although the coefficients $\Delta$ and $r_s$ are.  
Outside the virial radius $r_v$, the density profile must be
truncated so that the interior matter can be blended into the
background matter.  Typically the concentration
parameter $r_v/r_s$ exceeds $5$ for the present-day virialized objects,
and the $r^{-3}$ envelope in the density profile is noticeable. 
Such an universal profile has been tested against observations favorably
\cite{10}.

For our purposes, the universal profile can be extrapolated to describe
the configurations of future bound objects.  We will be
concerned about how the concentration parameter $r_v/r_s$ and the
characteristic density $\Delta$ vary with the
bound-object formation time.  Unfortunately, the virial radius $r_v$
is undefined in
Eq.(\ref{univ}), therefore prohibiting us from using this density profile
directly.  One hence needs to seek another method to address the
evolution of the universal profile.

The following strategy may be adopted for a quantitative
evaluation of the evolution of the profile.
The steady-state collisionless dark matter
is described by the phase-space
distribution function
$f(C_1({\bf x},{\bf p}), C_2({\bf x},{\bf p}), ...)$,
where $C_i$'s are the constants of motion.  For convenience,
we assume the momentum ${\bf p}$ to be isotropic and the mass density
spherically symmetric.  It follows $f(C_i)=f(E)$, a function of
only particle energy $E$.  Set $f(E)=0$ for unbound particles that 
have $E>0$.
Upon integrating out the momentum from $f(E)(=f(p^2/2m+\phi))$,
we obtain the mass density $\rho(\phi)$, a function of the
potential $\phi$ of all forces.  The universal profile 
Eq.(\ref{univ}) is obtained from the Poisson equation for an
appropriate $\rho(\phi)$.  The functional form of $\rho(\phi)$ 
can be approximated by the following considerations. 
The inner singular
density profile requires $\rho(\phi)\sim (\phi-\phi_{min})$ near
the potential minimum $\phi_{min}$ at $r=0$.  On the other hand,
the outer $r^{-3}$
behavior requires $\rho(\phi)\sim -\phi^3$ at the edge of the
potential well, $\phi\to 0$, where
the bound particles are about to escape.  
Thus these desired behaviors can be captured by
\begin{equation}
\rho(\phi)=-\frac{\alpha\phi^3}{\exp[\beta(\phi-\phi_{min})]-1}.
\label{den}
\end{equation}

Normalizing all quantities
in dimensionless forms, we arrive at the Poisson equation,
\begin{equation}
\frac{1}{x^2}\frac{d}{dx}(x^2\frac{d\psi}{dx}) + 1
= -\frac{g\psi^3} {\exp[b(\psi+1)]-1},
\label{pois}
\end{equation}
where $\psi\equiv \phi/|\phi_{min}|(\le 0)$,
$x\equiv r/L$ with the characteristic length
$L\equiv\sqrt{|\phi_{min}|/8\pi G\rho_\Lambda}$,
$b\equiv \beta|\phi_{min}|$ and
$g\equiv \alpha|\phi_{min}|^3/2\rho_\Lambda$.
The gravitational potential $\phi_g$
on the left-hand side of the Poisson equation has been replaced by
$\phi+(4\pi G/3)\rho_\Lambda r^2$.

The solutions to eq.(\ref{pois}) require numerical integration.
Different values of $g$ and $b$ yield different $\Delta$ and
$r_s$ of eq.(\ref{univ}).  This is a nonlinear eigenvalue problem,
and there exists a locus in the eigenvalue $(g,b)$ space,
along which the bound objects satisfy the desired boundary condition
that $\rho$ vanishes at the potential maximum.
Figure (1) presents the parameter $b$ as a function of $g$ for
these bound objects.  It is instructive to seek a measure for the
the inner $r^{-1}$ core region. 
Also plotted in Fig.(1) is the dimensionless half-mass-weighted
radius $h\equiv\langle x D\rangle_{1/2}$ of these virialized objects,
where $D\equiv\rho/2\rho_\Lambda$.
The quantity
$\langle x D\rangle_{1/2}\equiv 3\int_0^{x_{1/2}} x^3 D dx/x_{1/2}^3$,
and $x_{1/2}$ is defined to satisfy
$\int_0^{x_{1/2}} x^3D dx\equiv (1/2)\int_0^{x_{max}} x^3 D dx$
with $D(x_{max})=0$.  Note that the
parameter $g$ varies by a factor of $10^5$ and $b$
by several hundred,
and yet the quantity $h$ does not change in most parameter space
until near the high end of $(g,b)$, where it decreases only by $50\%$.
This result shows that the $r^{-1}$ core contains more than half of
the mass for most solutions, except for those with $g \ge 5\times 10^4$.

Plotted in Fig.(2) are the density and potential profiles of three
solutions with $(g,b)=(12, 0.0063),(5\times 10^4, 2)$, and
$(8\times 10^5, 3.2)$.  These solutions, denoted as
$(a), (b)$ and $(c)$ respectively, have
progressively stronger cores of smaller sizes as both $g$ and $b$ increase.
Solution (a), having an extended $r^{-1}$
core, gives no detectable outer envelope in the density profile, 
whereas solution (c), with a strong and
compact core, contains a distinct $r^{-3}$ envelope.
These solutions are expected to have progressively larger
ratios of $\rho_v/\rho_\Lambda$, corresponding to virialized objects formed
at progressively earlier epochs.

As virialized objects formed at different epochs appear differently,
a future observer ought to be able to date these objects by morphology.
To determine the formation epochs, we first start from the extended-core
solution (a), which is at the small end of the $(g,b)$ spectrum shown in
Fig.(1). It has already been well within a self-similar scaling regime
beginning around $(g,b)\sim (196, 0.1)$, where the denominator
of $\rho(\phi)$ in Eq.(\ref{den}) can be expanded to the leading order
of small $\beta$, yielding
$g/b\to 1953$.  Different solutions in the scaling
regime have the same profiles and hence
solution (a) can be identified to be
the virialized configuration of the last collapsed objects.
Making use of the result derived earlier
that $\rho_v/\rho_\Lambda=41$ within the virial radius $x_v$
for the last bound object, we
determine the mass fraction, equal to $0.972$, within the virial radius.

Assuming that the mass fractions, $0.972$,
within the virial radii of all solutions remain the same, we find that
$\rho_v/\rho_\Lambda= 65$ and $85$ for solutions (b) and (c),
respectively.
According to the virialization model analyzed earlier,
solutions (b) and (c) correspond to the virialized objects with
turn-around times at $z=0$ and $z=0.23$, or cases ii) and iii) 
considered earlier, and they
are to complete the collapse in the future
when the scaling factors $a/a_0$ become $2.25$ and $1.63$, respectively.

We note that the redshift evolution of morphology can be rather rapid
for these future virialized objects formed in between $a/a_0\sim 1.5$ and
$4.5$; by contrast,
the collapsed objects up to the present do not show much difference
in morphology, all having a compact core and a $r^{-3}$ envelope.

Common to all virialized objects shown in Fig.(2) is that
matter is mostly confined
at a considerable distance away from the vacuum radius $x_{max}$.
Moreover,
as the potential $\psi$ vanishes quadratically $(x-x_{max})^2$
near the potential maximum, the density should vanish as
$(x-x_{max})^6$, according to eq.(\ref{den}).     
Using the force balance of
pressure against gravity, it means that 
the local temperature vanishes as $(x-x_{max})^2$ at the edge.
That is, the dark matter is cold at the outer edge of the mini-universe.  
Moreover, conservation of
the phase-space volume for collisionless particles demands that
$\Delta r^3\Delta v^3$ remains a constant, and when $\Delta v^3\to 0$
we have $\Delta r\to\infty$. 
That is, the mini-universe contains a thick outer layer of cold
dark matter particles.  

How do these cold particles result from?
Some fraction of collapsed dark-matter particles
must have been pulled away by the cosmic repulsive force
to the background during virialization.  Thus, evaporation
should serve as an important
cooling mechanism to produce a thick outer layer of cold particles.
The mass loss due to evaporation may be estimated.
We note from Fig.(2) that the cool matter outside the virial radius $r_v$
occupies about $90\%$ of the volume enclosed by $x_{max}$, but it
amounts to only $2.8\%$ of the total mass.  The evaporated material can
not be much more than the cool matter located in the thick outer layer.
Therefore the small mass loss
should not create any problem for the above quantitative 
analyses that assume mass
conservation during virialization.

Finally, the typical masses of bound objects formed at various
epochs are also of relevance toward understanding the future world.
As the linear power spectrum $P(k)$ is well
understood\cite{11,12}, this issue can be properly addressed as well.
With a given conserved $\kappa_i$, the radius $r_i$ can be traced back to a
sufficiently high $z$ through Eqs.(\ref{mini}) and (\ref{freid}), 
thereby fixing the amplitude of linear
over-density by the relation $\delta\rho(z)=3\kappa_i/8\pi G r_i^2(z)$.
On the other hand, $\delta\rho(z)$ is also given by 
$g(z)\int_0^{\pi/r_i(z)}4\pi k^2 P(k) dk$, where $g(z)$ is the growth
factor of linear perturbations\cite{13}. Identification of the two
$\delta\rho$'s allows $r_i(z)$ to be determined, and
the enclosed mass is simply $M=4\pi\rho(z) r_i^3(z)/3$,
a redshift-independent quantity, as it should.
The typical mass of the last bound objects
(solution (a) of Fig.(2)) is found to be $2\times 10^{14}$ solar masses,
the mass of a poor galaxy cluster, and those for solution (b) and (c)
are $10^{14}$ and $4.5\times 10^{13}$ solar masses, 
respectively (assuming $H_0=65$ km/sec/Mpc).

\section{Bound Object Formation with Quintessence}

In this section we study the condition for bound object formation under
the repulsive force of the quintessence with a constant $w_q$. At present
experimental data allow for the possibility that the accelerating 
cosmic expansion
be due to quintessence with a relatively wide range of constant 
$w_q(\geq -1)$ and energy densities\cite{bond}.  

In order to have an accelerating universe at present, 
the equation-of-state parameter $w_q$ should satisfy
\begin{eqnarray}
w_q < -{1\over 3} (1+{\Omega_m^0\over \Omega_q^0}).
\label{wq}
\end{eqnarray}
Unfortunately, the value of $\Omega_m^0/\Omega_q^0$
is still rather uncertain, and its plausible value deduced from data
correlates with $w_q$, with a larger $\Omega_m^0/\Omega_q^0$
for a smaller $w_q$ in a flat universe. 
For example, 
when $\Omega_m^0 = 35\%$ and $\Omega_q^0 = 65\%$ to make up
the energy budget for a flat universe,
eq.(\ref{wq}) implies $w_q < -0.5$, but the existing data pushes the 
plausible $w_q$ to be near $-1$.
Nevertheless, to have an idea as to how the results vary with $w_q$,
we shall consider three representative values, $w_q = -0.5, -0.75$ and $-1$,
in our later discussions. 

The quintessence energy density varies with the scaling factor as
$\rho_q = \rho_q^0 (a/a_0)^{-3(1+w_q)}$. The reference background
cosmological time $t_B$ in this case is given by
\begin{eqnarray}
t_B = {1\over \sqrt{6\pi G \rho_q^0}}\int^1_0{dy\over \sqrt{\Omega_m^0/
\Omega_q^0 + y^{-2w_q}}}.
\label{time}
\end{eqnarray}

For the study of bound mini-universes, the 
situation becomes more complicated than that with a cosmological constant
because the space curvature $\kappa_i$ within the collapse region is
time-dependent, and eq.(4) is no longer valid.  
(The analysis presented in Ref.\cite{qq}, which assumes a constant $\kappa_i$, 
is hence invalid.)  This problem should be analyzed using the momentum
component of the Einstein equation, which describes the force balance of
inertia with gravity and quintessence force.
It can be understood as follows.  Since the self-gravity of quintessence is
negligible, the repulsive
force of quintessence at a radius $r$ varies with time as 
$(a_0/a(t))^{3(1+w_q)}$.  
This is a time-dependent force, thereby leading to a time-dependent space curvature.

The momentum component of the Einstein equation is a second-order 
differential equation: 
\begin{eqnarray}
{\ddot{r}\over r} = -{4\pi G\over 3} [\rho_i + (1+3w_q) \rho_q].
\end{eqnarray}
Defining $x= (a/a_0)$, $y=(r/r_0)$,  
one has
\begin{eqnarray}
&&\dot{x} = 
\sqrt{{8\pi G\over 3}\rho^0_q}\sqrt{(\Omega^0_m/\Omega^0_q)(1/x) + 
x^{-(1+3w_q)}},\nonumber\\
&&\ddot{y} = -{4\pi G\over 3}\rho^0_q
y[{\Omega_i^0\over \Omega^0_q} y^{-3} + (1+3w_q)x^{-3(1+w_q)}].
\label{force}
\end{eqnarray}
Combining these two equations one arrives at
\begin{eqnarray}
2\bar y'' \bar x (1 + \bar x^{-3w_q})
&=&\bar y'(1 +(1+3w_q)\bar x^{-3w_q})\nonumber\\
&-&( {\bar x^2\over \bar y^2}
+ (1+3w_q) {\bar y\over \bar x} \bar x^{-3w_q}),
\end{eqnarray}
where $\bar x\equiv x(\Omega_m^0/\Omega_q^0)^{1/3w_q}$,
$\bar y\equiv y(\Omega_m^0/\Omega_q^0)^{1/3w_q}(\Omega^0_m/\Omega_i^0)$ and
$\bar y'=d\bar y/d\bar x$.

We may solve eq.(15) to obtain the conditions for bound object
formation at any background cosmic time $t_B$, c.f., eq.(\ref{time}).  
However, due to the wide range of possible
parameters for the quintessence cosmology,
we shall instead concentrate on the critically bound mini-universe. 
The asymptotic state of the critical mini-universe can be found by 
letting both $\bar y\to\infty$ and $\bar x\to\infty$ 
in the infinite future.  The existence of such a state also requires 
that the background expansion be accelerating, i.e., $w_q<-1/3$.   
It then follows
that eq.(15) has a scaling solution, 
$\bar y=(\Omega_{i\infty}/\Omega_{q\infty})^{1/3}\bar x^{(1+w_q)}$, where
\begin{eqnarray}
{\Omega_{i\infty}\over\Omega_{q\infty}} = w_q^2-w_q,
\label{constraint}
\end{eqnarray}
by matching the coefficients in eq.(15). The subscript "$\infty$"
refers to evaluation of a quantity at a reference epoch in the infinite
future.  

The space curvature $\kappa_i$ for this mini-universe 
defined as minus of twice the sum of kinetic energy and 
potential energy (of the force used in eq.(\ref{force})) is given, 
in the asymptotic limit, by
\begin{equation}
\kappa_i=-({a\over a_\infty})^{-(1+w_q)}(1+3w_q)
{4\pi G \rho_{q\infty}r_\infty^2},
\label{kk}
\end{equation} 
where eq.(\ref{constraint}) has been used. This confirms the
statement made earlier that with quintessence the curvature is
time dependent.

The space curvature decreases with time as $a^{-(1+w_q)}$ but
its magnitude cannot be determined from the scaling solution;
it can be fixed only by integrating eq.(15) to obtain the entire
solution. 
In this asymptotic state, the absolute values of 
kinetic, potential and total energies all have the same time
dependence with comparable magnitudes.
For $w_q=-1$, we recover the cosmological constant case with
$\rho_{i\infty}/\rho_{q\infty} = 2$ and $\kappa_i$
given in eq.(\ref{cur}).
The above critical mini-universe is very different from that of
the case with a cosmological constant, in that its size grows 
indefinitely since $r$ grows as $a^{(1+w_q)}$. 

One may further extrapolate the ratio 
$\rho_{i\infty}/\rho_{q\infty}$ 
back to the present by integrating eq.(15)
numerically.  The results are shown in Fig. 3. In Fig. 3 
we show $\mbox{log}(\Omega_i/\Omega_q)$ as a function of 
$\mbox{log}(\Omega_q/\Omega_m)$ for
$w_q=-0.5, -0.75$ and $-1$. 
From Fig.3, one 
can read off the densities of the critically bound 
mini-universe on the vertical axis for various background densities on
the horizontal axis.  (The horizontal axis can be regarded as another way
for expressing the background cosmic time, c.f., eq.(\ref{time}).) 
Regions whose present local densities exceed 
by a finite amount above a given curve shown in Fig. 3
will turn around within a finite time and become virialized in a 
manner similar to that discussed in Sec.(2), though the details 
differ\cite{sw}. 

However, it is interesting to note that
there are differences for the virialized objects between the general 
quintessence cosmology and the $\Lambda$CDM cosmology.
For the case with quintessence, 
a local region with an energy infinitesimally smaller than the
critically-bound condition (local matter density larger than the critical
density by an infinitesimal amount), the region will undergo collapse in the
infinite future.  The bound object has
a nearly zero binding energy, as can be seen from eq. (\ref{kk}), 
and hence after virialization its virial
radius $r_v$ discussed in Sec.(2) will be infinitely large, a
great contrast to the $w_q=-1$ case where the just bound objects have
finite sizes.

As has been mentioned earlier in this section, the precise value of
$\Omega_m^0/\Omega_q^0$ is yet to be determined and the plausible value
of $\Omega_m^0/\Omega_q^0$ in fact anti-correlates with the value of $w_q$.  
Though for $w_q=-1$ the most plausible 
$\Omega_m^0/\Omega_q^0=0.35/0.65$, for $w_q=-0.75$ and $-0.5$ the most plausible
$\Omega_m^0/\Omega_q^0$ decreases to approximately $0.1/0.9$ and $(\approx)0/1$,
respectively\cite{bond}.
Precise knowledge for the content of energy forms,
which can be acquired from more accurate calibration of
Type Ia supernovae and measurements CMB radiation as well as other 
observations\cite{bond,snap}, are crucial for making good use of Fig.3.

\section{Discussions and Conclusions}

In this work, we have quantitatively shown that mini-universes can
form against the repulsive
force of cosmological constant or quintessence with a constant 
equation-of-state parameter $w_q$,
as long as they have sufficiently high densities at present.
These mini-universes, when virialized,
have distinctly different configurations depending on the formation time.
Bound objects form earlier appear more compact and are confined
by deeper gravitational potentials.  This issue is addressed 
in details for the $\Lambda$CDM cosmology, where
the mini-universes are found typically no heavier than a poor galaxy
cluster.  
 
As discussed in Sec.(2), the actual size of mini-universe varies, depending
on the primordial density power spectrum.  It is instructive to examine mini-universes
of very large scale.  We note that
a marginally bound mini-universe has a  
gravitational radius
$r^2_g = c^2r^2_{max}/\kappa_i =c^2/(3\Omega_\Lambda H_0^2)$ in the case of 
$\Lambda$-cosmology.  If a 
marginally bound mini-universe has a physical size larger than $r_g$, 
no light emitted within can escape from it, and hence
behaves like a black hole when 
observed from the exterior.  Such a
mini-universe is
absolutely stable against the background pulling force. 
We therefore obtain 
the minimal size $r_{min}$ of an absolutely 
stable and marginally bound mini-universe 
to be $c/\sqrt{3\Omega_\Lambda} H_0$, which is comparable to the Hubble 
radius of the background universe.  

For mini-universes with larger
local densities, the required radius is smaller.  These mini-universes
will eventually undergo catastrophic collapse, and can
asymptotically be described by a static space-time metric --
the Schwartzschild-de Sitter metric:
\begin{equation}
ds^2= (1-{2GM\over r} - {\Lambda\over 3}r^2)d^2t
- ( {d^2r\over 1-2GM/r - \Lambda r^2/3} + 
r^2 d^2\theta + r^2 \sin^2\theta d^2\phi).
\end{equation}  
It has two horizons: an inner horizon at a modified Schwartzschild radius 
and an outer horizon pertaining to the accelerating cosmic expansion.  
The region in between the two horizons is unstable, in that an observer located
in between the two horizons must eventually fall onto either of them.

Our local universe cannot be
within such a closed universe, as the angular scale of the 
first peak in the CMB anisotropy power spectrum indicates the local universe
within the Hubble radius to be spatially flat\cite{7a}.  This, however, does not
preclude the possibility that regions outside the present Hubble horizon may have 
sufficiently large super-horizon over-densities to become a closed universe.  
The required present
matter over-density $\Omega_i^0/\Omega_m^0$ in such universes is as high as $6.75$.  
These regions are very rare if the primordial density fluctuation is indeed 
scale-invariant, created during the inflation in early universe,
with a power spectrum linearly proportional to the wavenumber $k$.  Such local
closed universes should be more abundant in the past, and some of them may have 
been within our present horizon, perhaps in the form of super-massive black holes.

In the case of quintessence,
the situation is more complicated because the mini-universe has
a time-dependent space curvature.  The curvature actually decreases with time,  
due to the fact that matter and quintessence as a whole can be regarded as a
non-ideal fluid and 
the local matter over-density gives rise to entropy
fluctuations.  Space curvature fluctuations pertinent to entropy fluctuations 
always decrease monotonically in time, regardless whether the fluctuations
are of sub-horizon size or super-horizon size.  However, adiabatic
super-horizon fluctuations can also give rise to 
curvature fluctuations with no time dependence, 
for which quintessence fluctuations can grow and collapse as well.
They are more relevant for the discussions of the
local closed universe in the context of quintessence cosmology.   
Unfortunately, our analysis presented in Sec.(3) is confined to
entropy fluctuations, where the quintessence fluctuation is 
negligible.  Nevertheless, the dynamics of such a local closed 
universe with a constant space curvature is relatively simple and 
like that of a homogeneous universe.  
Thus the discussions given in the preceding paragraph for $\Lambda$-cosmology
also holds for quintessence-cosmology.  The super-massive black hole
so formed should contain a substantial amount of dense quintessence field within
the Schwartzschild radius, and its asymptotic space-time metric depends on the
yet unknown nature of the quintessence field.  

Return to the mini-universes arising from entropy fluctuations and 
of sub-horizon size analyzed in Sec.(3). 
As long as the matter density is larger than the
critical density given in Fig. 3 for the corresponding equation of state
parameter, the local region will be dominated by the matter gravity 
and collapses to become a mini-universe.
These mini-universes will become virialized in a similar 
manner as those of the $\Lambda$CDM cosmology\cite{sw}, due to that 
virialization depends mainly on the balance between gravity and pressure,
and to a lesser degree on the large-scale cosmic pulling force.  Despite
the similarity, the details can be different.  
For example, conservation of total energy 
within the bound object, which is essential for applying the virial condition,
no longer holds in quintessence-cosmology, and hence the resulting virial
over-density within the object deviates from that of the $\Lambda$CDM cosmology.
Also the marginally bound mini-universe has an infinitely large virial radius.  

At any rate,
a single tightly bound mini-universe will, in the future, 
emerge from the Local Group, whose
appearance resembles solution (c) of Fig.(2)
in the case of $\Lambda$CDM cosmology. 
Interior to the mini-universe, local physics in the future will remain
the same as what we experience presently.  However,
the mini-universes will become isolated island universes in the
future where
no communication among them is possible, and each is a lone
world.  When observed within a given mini-universe,
all other mini-universes originally within the horizon will
eventually fall onto the horizon and become frozen.  Lights emitted from
the horizon are frozen as well, making other island-universes
dark and invisible.

The detailed properties of the mini-universes, 
depend on whether the
accelerating expansion is due to the cosmological constant or quintessence.
New experimental data from various observations in the coming decades are 
expected to reveal more information.   By then we will have a better 
understanding of the properties of mini-universes.  
If the acceleration of our universe at present is indeed due to 
cosmological constant or quintessence with constant $w_q$, 
our universe will enter a new era, the era where all 
island universes are falling onto each other's horizon and appear to 
fade away.  It is, nevertheless, comforting to know that the Milky
Way galaxy is in one of these island universes, within each of which most 
physics remains the same as is today.  Review on the details of how 
an island universe will eventually end itself 
with physics known today can be found in Ref.[19].

This work is supported in part
by National Science Council under grants NSC 89-2112-M-002-058 and NSC
89-2112-M-002-065, and in part
by the Ministry of Education Academic Excellence Project 89-N-FA01-1-4-3.

\begin{figure}[htb]
\centerline{\DESepsf(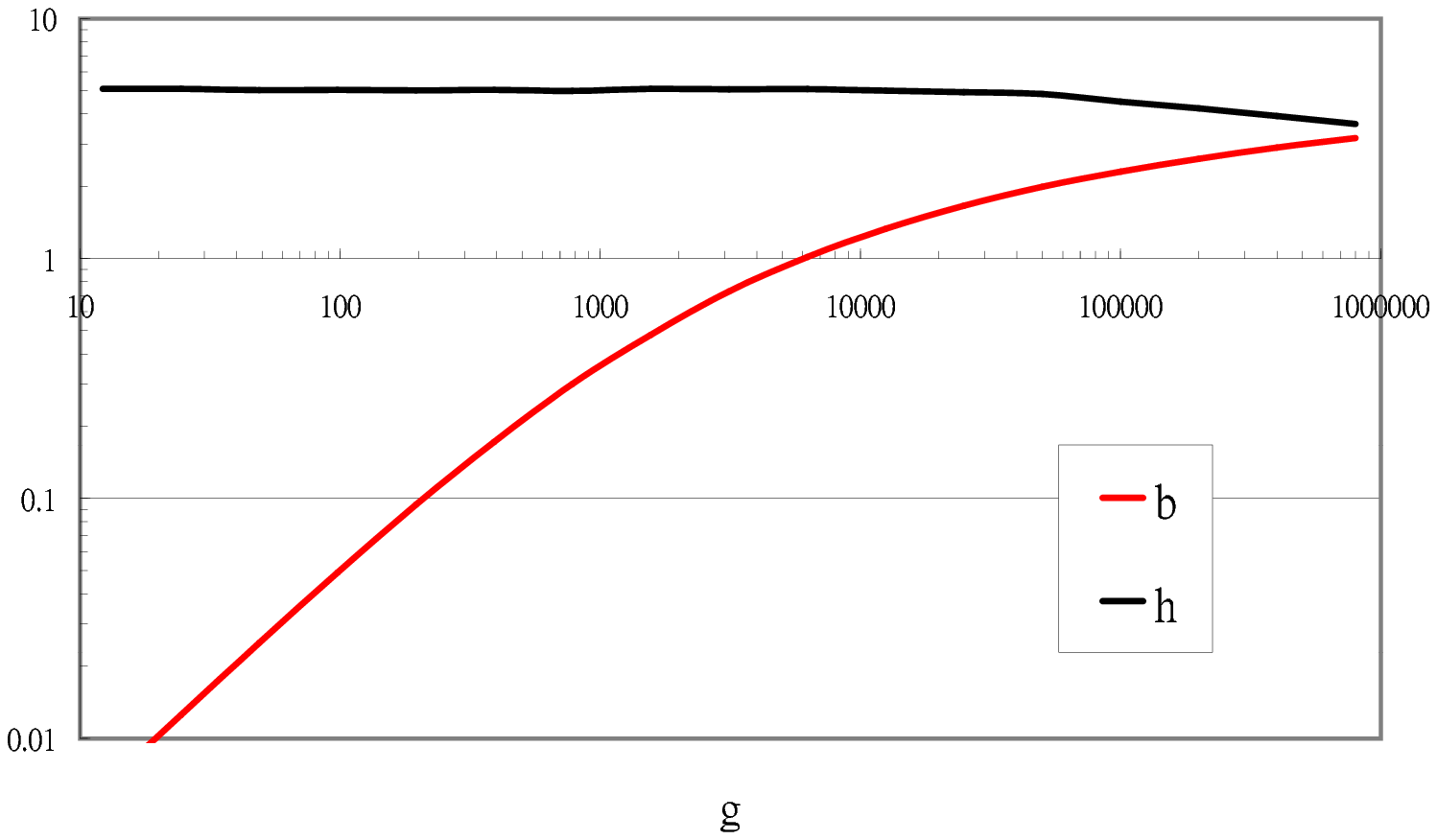 width 12cm)}
\smallskip
\caption{Nonlinear eigenvalues $(g,b)$ for Eq.(\ref{pois}).  
When $b<0.1$, the solutions are in a similarity scaling regime.  
Also shown is the half-mass weighted radius, $h$.  
The constancy of $h$ implies that
the $r^{-1}$ core contains more than half of the total mass.}
\bigskip\bigskip\bigskip\bigskip\bigskip
\centerline{\DESepsf(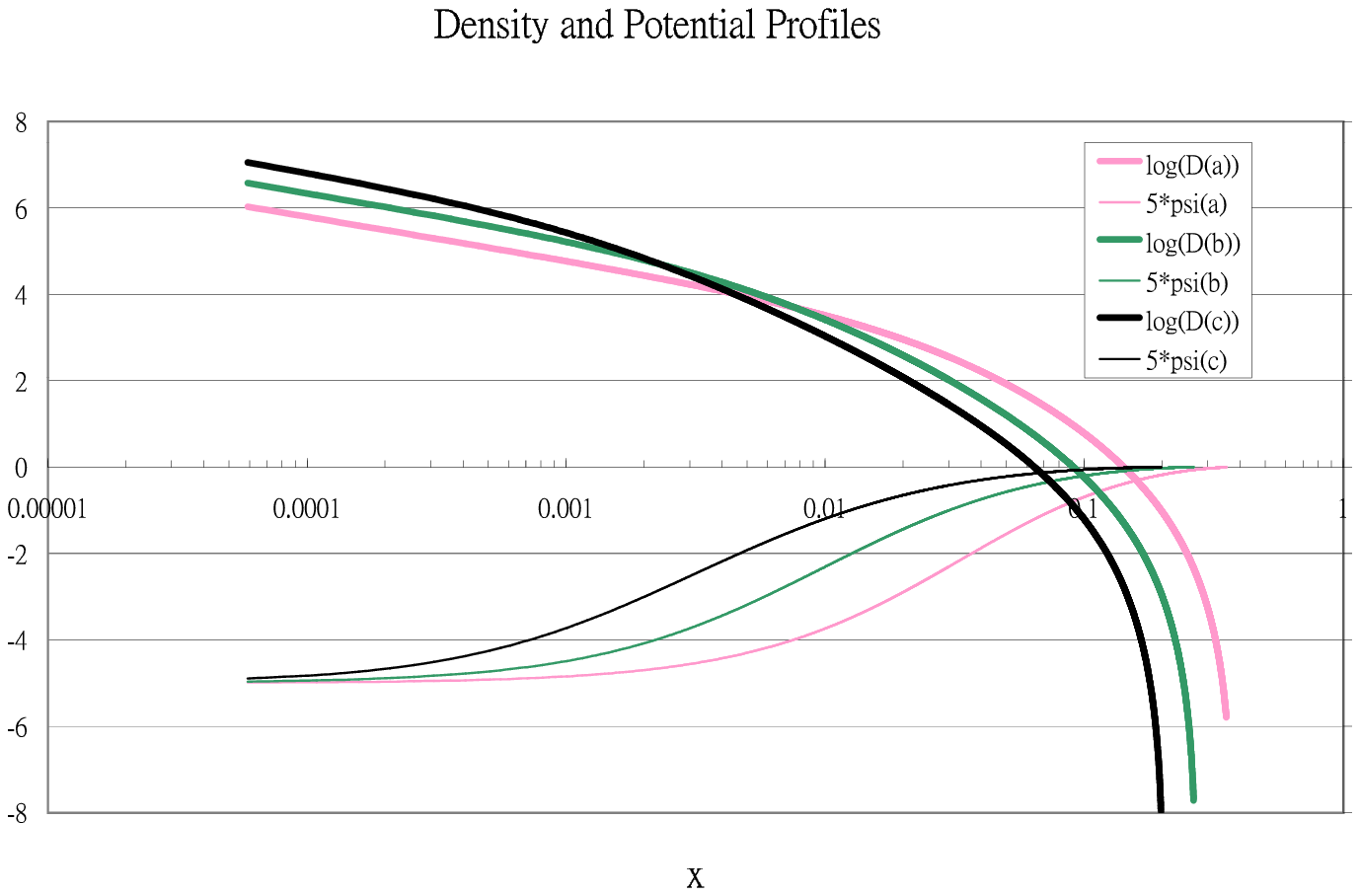 width 12cm)}
\smallskip
\caption{Dimensionless densities $D$ (in $log$ scale) and potentials
$\psi$ for solutions $(a), (b)$ and $(c)$ of Eq.(\ref{pois}).}

\centerline{\DESepsf(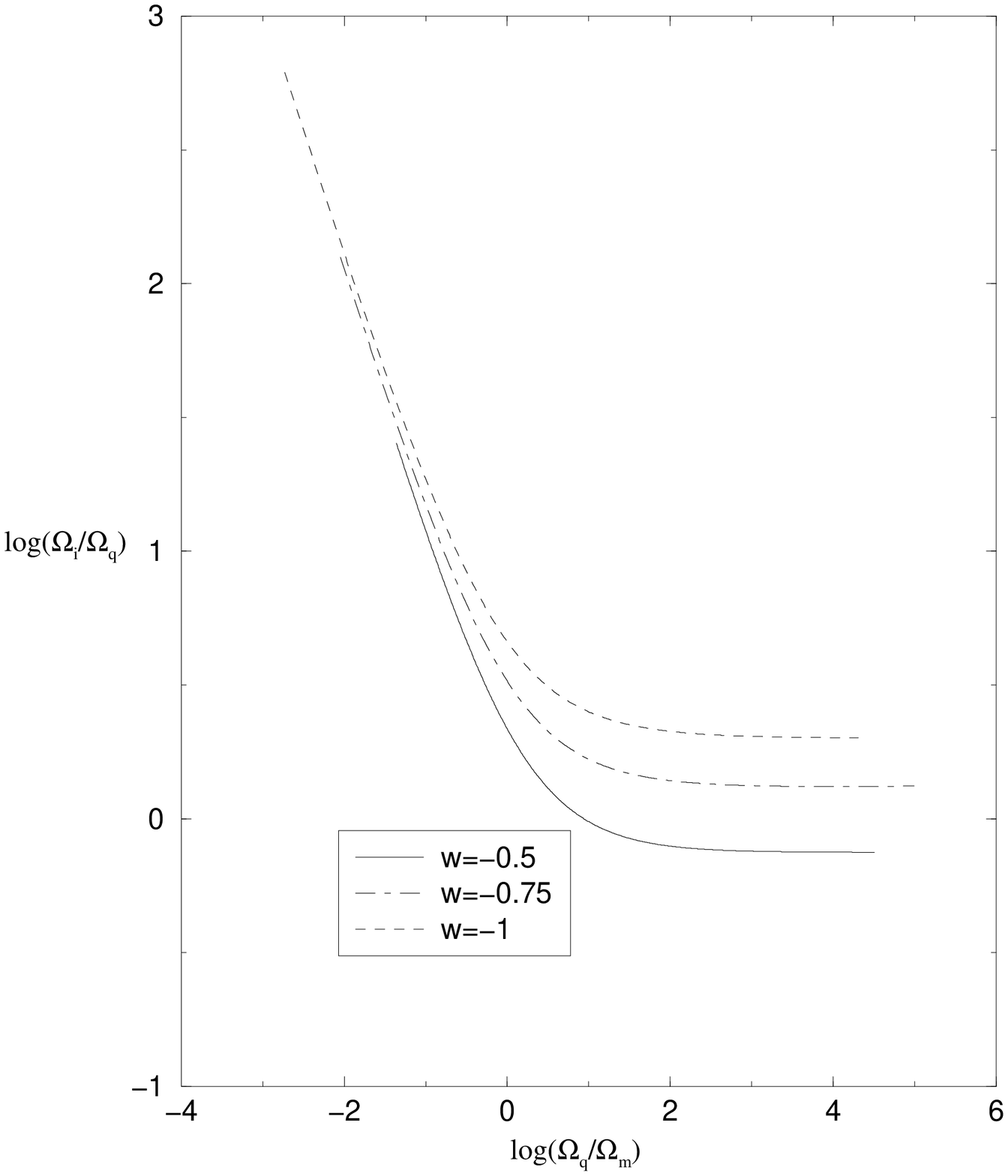 width 8cm)}
\vspace{1.5cm}
\caption{
The barely bound mini-universe critical density as a function of time.
The vertical axis is $\mbox{log}(\Omega_i/\Omega_q)$ and the horizontal axis is 
$\mbox{log}(\Omega_q/\Omega_m)$.}
\bigskip\bigskip\bigskip\bigskip\bigskip
\end{figure}


\begin{thebibliography}{99}
%

\bibitem{1} S. Perlmutter et al., Nature {\bf 391}, 51(1998); Astrophys. J.
{\bf 517}, 565(1999); A.G. Riess et al., Astron. J. {\bf 116}, 1009(1998).

\bibitem{2} S. Hawking and G. Ellis, The Large Scale Structure of Space-time,
Combridge University Press, 1973.

\bibitem{3} J. Barrow, R. Bean and J. Magueijo,
Mon. Not. Astron. Soc. 316, L41-4(2000);
A. Albrecht and C. Skordis, Phys. Rev. Lett. {bf 84}, 2076(2000);
S.M. Carroll, Phys. Rev. Lett. {\bf 81}, 3067(1998); E. Halyo e-print
hep-ph/0105216; J. Cline, e-print hep-ph/0105251.

\bibitem{4} A.G. Riess et al., e-print astro-ph/0104455.

\bibitem{5} S. Hellerman, N. Kaloper and L. Susskind, e-print hep-th/0104180;
W. Fischler et al., e-print hep-th/0104181.

\bibitem{6} X.-G. He, e-print astro-ph/0105005; 
J. Moffat, e-print hep-th/0105017;
J.-A. Gu and W.-Y, Hwang, e-print astro-ph/0106387.

\bibitem{aa}
R.R. Caldwell, R. Dave, and P.J. Steinhardt,
Phys.\ Rev.\ Lett.\  {\bf 80}, 1582 (1998) 
I.~Zlatev, L.~Wang and P.J.~Steinhardt,
Phys.\ Rev.\ Lett.\  {\bf 82}, 896 (1999) 
P.J.~Steinhardt, L.~Wang and I.~Zlatev, 
Phys.\ Rev.\ D {\bf 59}, 123504 (1999).

\bibitem{bb}
Y.\ Fujii, Phys.\ Rev.	{\bf D26}, 2580 (1982); L.H.\ Ford,
 Phys.\ Rev.  {\bf D35}, 2339 (1987); P.J.~Peebles and B.~Ratra,
Astrophys.\ J.\  {\bf 325}, L17 (1988);  Y.\ Fujii and T.\ Nishioka,
Phys.\
 Rev.\ {\bf D42}, 361 (1990); K.\ Sato, N.\ Terasawa, and J.\
Yokoyama,
in Proc.\ XXIVth Recontre de Moriond, 
``The Quest for the Fundamental
Constants in Cosmology,'' eds.\ J.\ Audouze and J.\ Tran Thanh Van.
(Editions Fronti\`{e}res, France 1990) 193;  T.\ Nishioka and S.\
Wada, 
Int.\ J.\ Mod.\ Phys.  {\bf A8}, 3933 (1993).

\bibitem{cc} N.\ Arkani-Hamed, L.J.\ Hall, C.\ Kolda, H.\
Murayama,
Phys.\ Rev.\ Lett. {\bf 85}, 4434 (2000);
S.M.\ Barr and	D.\ Seckel,  hep-ph/0106239.

\bibitem{7a} P. de Bernardis et al., Boomerang Coll., Nature {\bf 404},
955(2000); S. Hanany et al., Maxima
Coll., Astrophys. J. {\bf 545}, L1-L4 (2000).

\bibitem{bond} See for example M. Signore and D. Puy, e-print astro-ph/0108515;
P. Binetray, e-print hep-ph/0005037.

\bibitem{8} J.A. Peacock, "Cosmological Physics", Cambridge Univ. Press
(1999); A.R. Liddle and D. Lyth, "Cosmological Inflation and
Large-Scale Structures", Cambridge Univ. Press (2000).

\bibitem{7} O. Lahav, P.B. Lilje, J.R. Primack and M.J. Rees,
Mon. Not. Roy. Astro. Soc. {\bf 251}, 128 (1991).


\bibitem{9} J.F. Navarro, C.S. Frenk and S.D.M. White, Astrophys. J.
{\bf 462}, 563 (1996); J.F. Navarro, C.S. Frenk and S.D.M. White,
Astrophys. J. {\bf 490}, 493 (1997).

\bibitem{10} J.A. Tyson, G.P. Kochanski and
I.P. dell'Antonio, Astrophys. J. {\bf 498}, L107 (1998); 
L.L.R. Williams,
J.R. Navarro and M. Bartelmann, Astrophys. J. {\bf 527}, 535 (1999).

\bibitem{11} G. Efstathiou, J.R. Bond and S.D.M. White,
Mon. Not. Roy. Astro. Soc., {\bf 258}, 1 (1992). 

\bibitem{12} C. Pryke, N.W. Halverson, E.M., Leitch, J. Kovac,
J.E. Carlstrom, W.L. Holzapfel and M. Dragovan, e-print astro-ph/01004490.

\bibitem{13} S.M. Carroll, W.H. Press and E.L. Turner,
Ann. Rev. Astron. Astrophys., {\bf 30}, 499 (1992).

\bibitem{qq} E. Lokas and Y. Hoffman, e-print astro-ph/0108283.

\bibitem{sw} L. Wang and P. Steinhardt, e-print astro-ph/9804015, Astrophys. 
J. 508, 483(1998). 

\bibitem{snap} P. Nugent, SNAP collaboration, in San Juan 2000,
Particle Physics and Cosmology, pp263 (2000).

\bibitem{adams} F. Adams and G. Laughlin, Rev. Mod. Phys. {\bf 69}, 337(1997).

\end{thebibliography}
\end{document}